\documentclass[10pt,twocolumn,letterpaper]{article}

\usepackage[pagenumbers]{iccv} 

\definecolor{iccvblue}{rgb}{0.21,0.49,0.74}
\usepackage[pagebackref,breaklinks,colorlinks,allcolors=iccvblue]{hyperref}

\def\paperID{*****} 
\def\confName{ICCV}
\def\confYear{2025}

\title{Advancing Lung Disease Diagnosis in 3D CT Scans}

\author{
 Qingqiu Li$^1$, Runtian Yuan$^1$, Junlin Hou$^2$, Jilan Xu$^1$, \\
Yuejie Zhang$^1$, Rui Feng$^1$, Hao Chen$^2$\\
\\
$^1$ Fudan University\\
$^2$ The Hong Kong University of Science and Technology\\
}

\begin{document}
\maketitle
\begin{abstract}
To enable more accurate diagnosis of lung disease in chest CT scans, we propose a straightforward yet effective model. Firstly, we analyze the characteristics of 3D CT scans and remove non-lung regions, which helps the model focus on lesion-related areas and reduces computational cost. We adopt ResNeSt50 as a strong feature extractor, and use a weighted cross-entropy loss to mitigate class imbalance, especially for the underrepresented squamous cell carcinoma category. Our model achieves a Macro F1 Score of 0.80 on the validation set of the Fair Disease Diagnosis Challenge, demonstrating its strong performance in distinguishing between different lung conditions.
\end{abstract}    
\section{Introduction}
\label{sec:intro}

\begin{figure}[t]
\centering
\includegraphics[width=\linewidth]{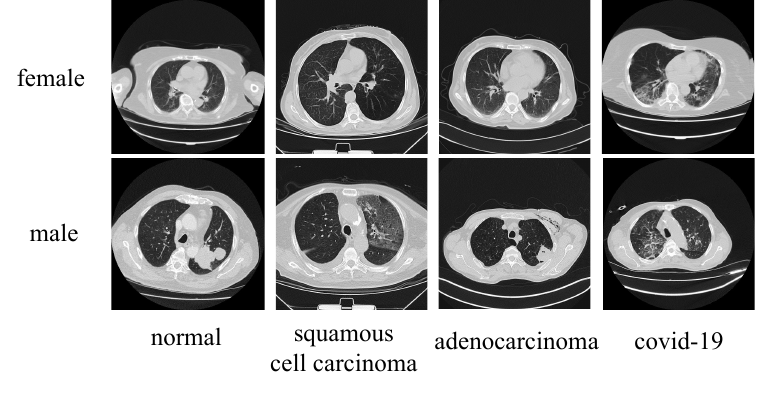}
\caption{Samples of different lung diseases.}
\label{fig:example}
\end{figure}

Lung disease remains a leading global health concern, with lung cancer and COVID-19 posing significant challenges to accurate diagnosis. Lung cancer, particularly non-small cell lung cancer, is one of the most prevalent and deadly malignancies worldwide. Among its subtypes, adenocarcinoma and squamous cell carcinoma are the most common. COVID-19, a respiratory illness, has also caused widespread health crises and fatalities in recent years. As illustrated in Figure.~\ref{fig:example}, chest CT scans have been widely used for diagnosing and monitoring affected patients, owing to their ability to provide detailed insights into the extent and severity of lung involvement. However, the large volume of CT images generated places a heavy burden on radiologists and medical professionals, making the diagnostic process increasingly challenging.

In recent years, deep learning has been widely applied to the automatic diagnosis of lung diseases using chest CT scans~\cite{arsenos2022large,arsenos2023data,gerogiannis2024covid,hou2021cmc,hou2021periphery,hou2022boosting,hou2022cmc_v2,kingma2014adam,kollias2018deep,kollias2020deep,kollias2020transparent,kollias2021mia,kollias2022ai,kollias2023ai,kollias2023deep,kollias2024domain,kollias2024sam2clip2sam,krueger2021vrex,li2024advancing,yuan2024domain}. However, previous methods often decompose 3D CT volumes into individual 2D slices for analysis or adopt relatively simple network architectures trained from scratch, resulting in suboptimal classification performance. To address these limitations, we treat each CT scan as a complete 3D volume and remove non-lung regions that do not contribute to disease identification. We then employ ResNeSt50~\cite{zhang2022resnest} as the backbone feature extractor and design a weighted cross-entropy loss function based on the proportion of lesion volume to address class imbalance.

Our main contributions are as follows:
\begin{enumerate}
\item We analyze the structural characteristics of 3D CT scans and remove irrelevant non-lung regions, enabling the model to focus on lesion-related areas and reducing computational overhead.
\item We utilize ResNeSt50 as a powerful feature extractor and introduce a lesion-aware weighted cross-entropy loss to mitigate the imbalance caused by underrepresented classes, i.e., squamous cell carcinoma.
\item Our model achieves a Macro F1 score of 0.80 on the validation set of the Fair Disease Diagnosis Challenge, demonstrating strong classification performance.
\end{enumerate}

\section{Methodology}
\label{sec:method}

\begin{figure*}[t]
\centering
\includegraphics[width=\linewidth]{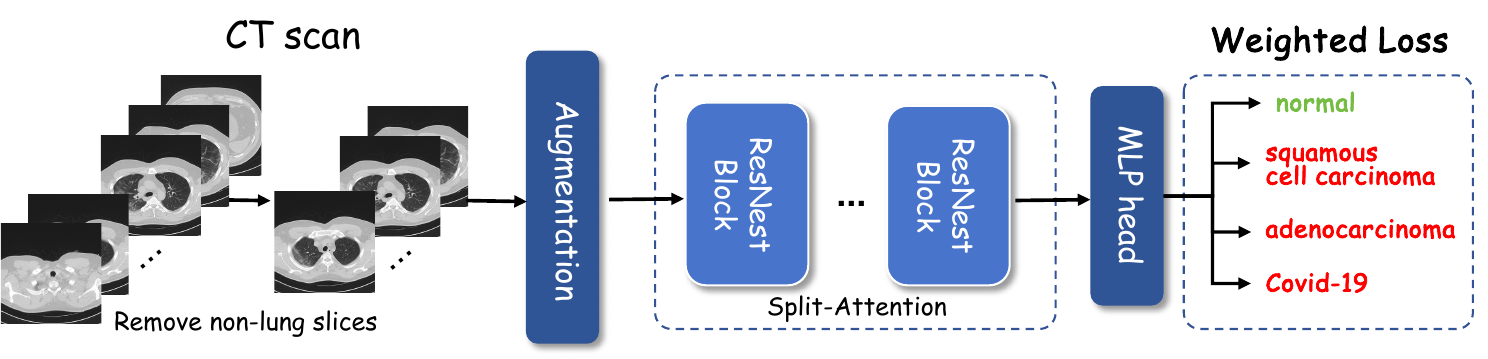}
\caption{Overview of our framework for lung disease diagnosis.}
\label{fig:framework}
\end{figure*}

The overall framework of our model is illustrated in Figure~\ref{fig:framework}. We begin by analyzing the 3D CT scans to identify slices that are irrelevant for disease detection, such as the upper neck region at the beginning and the abdominal area toward the end. These non-informative slices, which do not contain lung tissue, are removed. This allows the model to concentrate on regions that are more relevant to lesion detection and also helps reduce computational requirements.

After this preprocessing step, we employ ResNeSt50 as our feature extractor. This architecture includes a modular split attention mechanism within its network blocks, enabling the model to capture interactions across different feature-map groups and enhance representational power. To address the issue of class imbalance in the dataset, we use a weighted cross-entropy loss $\mathcal{L}_{\mathrm{wCE}}$:
\begin{equation}
\mathcal{L}_{\mathrm{wCE}} = - \sum_{c=1}^{C} w_c \cdot y_c \cdot \log(\hat{y}_c),
\end{equation}
where $w_c$ represents the weight of class $c$, and $y$ and $\hat{y}$ denote the ground truth and prediction for class $c$, respectively. 
This approach gives more weight to underrepresented classes, helping the model to learn more effectively from limited samples.

\section{Datasets}

\begin{table}[ht]
\centering
\resizebox{\linewidth}{!}{
\begin{tabular}{lccccc}
\hline
\textbf{Set} & \textbf{Total} & \textbf{A} & \textbf{Covid} & \textbf{G} & \textbf{Normal} \\
\hline\hline
Training & 330/404 & 125/125 & 100/100 & 5/79 & 100/100 \\
Validation & 78/77 & 25/25 & 20/20 & 13/12 & 20/20 \\
\hline
\end{tabular}
}
\caption{Statistics of the CT dataset, presented in the form of female/male.}
\label{tab:data}
\end{table}

We evaluate our proposed approach on the dataset provided by the Fair Disease Diagnosis Challenge. The dataset consists of chest CT scans from subjects diagnosed with lung cancer, COVID-19, or identified as healthy. Each scan is annotated to indicate whether it belongs to a healthy subject (Normal) or a subject diagnosed with Adenocarcinoma (A), Squamous Cell Carcinoma (G), or COVID-19 (Covid). Additionally, each scan includes information about the subject's gender (male or female). Detailed data statistics are summarized in Table~\ref{tab:data}.

\section{Experiments }
\subsection{Data Pre-Processing}
Our data pre-processing procedure is as follows. All 2D chest CT scan series are composed into a 3D volume of shape ($D$,$H$,$W$), where $D$,$H$,$W$ denotes the number of slice, height, and width, respectively. Then, each 3D volume is resized to dimensions of (64, 256, 256). Finally, we transform the CT volume to the interval [0, 1] for intensity normalization.

\subsection{Implementation Details}
We utilize 3D ResNeSt50 as the backbone of our model. For training, data augmentations include random resized cropping on the transverse plane, random cropping on the vertical section to 64, rotation, and color jittering. We use the Adam algorithm~\cite{kingma2014adam} as our optimizer, setting the learning rate to 1e-4 and the weight decay to 1e-5. Our model is trained for 100 epochs on 4 RTX 3090 GPUs with a batch size of 2 per GPU. Macro F1 score is used as the evaluation metric, which calculates the F1 score for each category separately and then averages these scores to assess overall performance.

\subsection{Experimental Results}
Our model achieves a Macro F1 Score of 0.80 on the validation set of the Fair Disease Diagnosis Challenge, demonstrating its strong performance in distinguishing between different lung conditions.

\section{Conclusion}
In this work, we present a simple yet effective approach for the classification of lung diseases from chest CT scans. By removing non-lung regions from 3D volumes, our model is able to concentrate on the most relevant anatomical structures, enhancing both efficiency and accuracy. Leveraging ResNeSt50 as a powerful feature extractor, combined with a weighted cross-entropy loss to address class imbalance, our method demonstrates strong diagnostic performance.
{
    \small
    \bibliographystyle{ieeenat_fullname}
    \bibliography{main}
}

\end{document}